\documentclass[prl,twocolumn,tightenlines,superscriptaddress,a4paper,lengthcheck]{revtex4-1}
\usepackage{amsfonts}
\usepackage{amsmath}
\usepackage{amssymb}
\usepackage{graphicx}
\usepackage{graphicx,amssymb,amsbsy,color}
\usepackage{bm}
\usepackage{bbold}

\begin{document}

\title{Thermalization and localization of an oscillating Bose-Einstein condensate in a disordered trap}
\author{Che-Hsiu Hsueh}
\affiliation{Department of Physics, National Taiwan Normal University, Taipei 11677, Taiwan}
\author{Russell Ong}
\affiliation{Department of Physics, National Taiwan Normal University, Taipei 11677, Taiwan}
\author{Jing-Fu Tseng}
\affiliation{Department of Physics, National Taiwan Normal University, Taipei 11677, Taiwan}
\author{Makoto Tsubota}
\email{tsubota@sci.osaka-cu.ac.jp}
\affiliation{Department of Physics, Osaka City University, Sugimoto 3-3-138, Sumiyoshi-ku, Osaka 558-8585, Japan}
\affiliation{The OCU Advanced Research Institute for Natural Science and Technology (OCARINA), Osaka, Japan}
\author{Wen-Chin Wu}
\email{wu@ntnu.edu.tw}
\affiliation{Department of Physics, National Taiwan Normal University, Taipei 11677, Taiwan}
\date{\today}

\begin{abstract}
We numerically simulate an oscillating Bose-Einstein condensate in a disordered trap
[Phys. Rev. A {\bf 82}, 033603 (2010)] and the results are in good agreement with the experiment.
It allows us to verify that total energy and particle number are conserved
in this quantum system. The disorder acts as a medium,
which results in a relaxation from nonequilibrium to equilibrium, {\em i.e.}, thermalization.
An algebraic localization is realized when the system approaches the
equilibrium, and if the system falls into the regime
when the healing length of the condensate exceeds the correlation
length of the disorder, exponential Anderson localization is to be observed.
\end{abstract}
\maketitle

Anderson localization (AL) had been a long-studied phenomenon in
electronic systems \cite{PhysRev.109.1492} . When transporting in
an environment with random disorder, waves of electrons
get localized after multiple scattering with the disorder. Recently there is a resurgence
of studies of AL in a variety of systems such as the photonic crystals
\cite{PhysRevLett.103.013901,NaturePhotonics.7.197}, the ultrasound in 3D elastic networks
\cite{NaturePhysics.4.945}, the quantum chaotic systems \cite{PhysRevLett.101.255702}, and the
cold atoms \cite{Nature.453.891,Nature.453.895}.
The experiment \cite{Nature.453.891} of cold atoms was done by expanding
Bose condensate in a weak random potential in which the healing length
($\xi$) of the condensate exceeds the correlation length ($\sigma_{\textrm{D}}$) of the
disorder, $\xi > \sigma_{\textrm{D}}$.
In this regime, the particle interaction is relatively unimportant to which
$\xi\rightarrow\infty$ corresponds to a noninteracting limit.
The experiment \cite{Nature.453.891} confirmed that the localized condensate exhibits an
{\em exponential} density profile
in a one-dimensional geometry, in agreement with the theory of Sanchez-Palencia {\em et al.}
\cite{PhysRevLett.98.210401}.

In this Letter, we show that an oscillating condensate in a disordered trap,
such as the experiment done by Dries {\em et al.} \cite{PhysRevA.82.033603},
can also exhibit AL when it comes to equilibrium and if it
falls into the regime $\xi > \sigma_{\textrm{D}}$.
Using exactly the same parameters of the experiment reported in
Fig.~2 of Ref.~\cite{PhysRevA.82.033603} for $\xi < \sigma_{\textrm{D}}$,
we perform a numerical simulation based on the Gross-Pitaveskii (GP)
approach in the presence of a spatially random disorder potential.
The results are in good agreement with the experiment as far as the overall
spatial dynamics is concerned. It allows us to verify that
when time passes a relaxation time $t_c$ (discussed later),
the system enters an {\em algebraical} localized
state \cite{PhysRevLett.98.210401}.
This motivates to carry out another simulation with the same parameters
except by reducing $\sigma_{\textrm{D}}$ to make $\xi > \sigma_{\textrm{D}}$.
In this case, exponential AL is eventually observed.

Another important factor in such a system is that it provides a simple framework
to investigate the long-standing question on how an ``isolated" many-body quantum
system, without coupling to the reservoir, can relax to a steady state that
seems to be in thermodynamic equilibrium, {\em i.e.}, thermalization
\cite{RevModPhys.83.863,doi:10.1146/annurev-conmatphys-031214-014726,0034-4885-79-5-056001}.
Our results show that total energy and particle number of the system are conserved
and temporal entropy reveals that a relaxation process from nonequilibrium to equilibrium
does exhibit. Random disorder plays the role of a {\em medium} (or {\em transistor})
which results in the exchange of
partial kinetic energy with partial potential energy.
When $t\gg t_c$, the system reaches a thermodynamic equilibrium in which
both kinetic and potential energies come to a constant.
There is no dissipation of the total energy through any kind of \emph{friction}.
It becomes evident that the thermalization to equilibrium is accompanied by
the localization in the system.

To make a direct comparison with the experiment reported in Fig.~2 of Ref.~\cite{PhysRevA.82.033603},
we consider a 1D Bose gas with a repulsive contact interaction that is trapped in a harmonic potential $V_{\textrm{ho}}(z)=m\omega^{2}z^{2}/2$.
In the dilute and ultracold condition, the condensate
wave function $\psi\left(z,t\right)$ is governed by the GP equation
in the presence of a real spatially random disordered potential
$V_{\textrm{dis}}(z)$,
\begin{equation}\label{GPE}
  i\hbar\partial_{t}\psi=\left[-\frac{\hbar^{2}}{2m}\partial_{z}^{2}+
  V_{\rm ho}(z)+V_{\textrm{dis}}(z)+Ng\left|\psi\right|^{2}-\mu\right]\psi.
\end{equation}
Here $N$ is the total number of atoms, $g$ is the coupling constant of contact interaction,
$\mu$ is the chemical potential, and $\psi$ is normalized to one, $\int|\psi|^{2}dz=1$.
The healing length at the center of the condensate is defined as $\xi=\hbar/\sqrt{2m\mu}$.
The disorder correlation length, $\sigma_{\textrm{D}}$, is defined by fitting the autocorrelation function $\langle V_{\textrm{dis}}(z)V_{\textrm{dis}}(z+\Delta z)\rangle=V_{\textrm{D}}^{2}\exp(-2{\Delta z}^{2}/\sigma_{\textrm{D}}^{2})$
with $V_\textrm{D}$ the strength of $V_{\textrm{dis}}(z)$ \cite{PhysRevA.77.033632}.

In the experiment, the condensate is released at a position off the center of the harmonic
trap that results in the subsequent oscillations.
For numerical convenience, we take an alternative scheme such that
the condensate is released at the trap center but with an initial velocity.
To obtain an initial wave function with a velocity $v_{0}$, we apply the
Galilean transformation: $\psi=\varphi\exp(imv_{0}z)$ and in the absence of
disorder, the corresponding GP equation for the residual wave function $\varphi$ is
\begin{equation}\label{GPE2}
  i\hbar\partial_{t}\varphi=\left[\frac{1}{2m}\left(\frac{\hbar}{i}\partial_{z}-mv_{0}\right)^{2}
  +V_{\textrm{ho}}\left(z\right)+Ng\left|\varphi\right|^{2}-\mu\right]\varphi.
\end{equation}
Long-term imaginary-time evolution of Eq.~(\ref{GPE2}) gives
$\varphi$ which in turn gives the initial wave function $\psi$.
A cutoff wavevector $k_c$ corresponding to
the shortest length scale or the largest $k$ scale in association with
the healing length $\xi$ is naturally introduced in the simulation which gives the best results.
In natural units: $\hbar=m=\omega=1$,
experimental parameters are $\mu=200$, $v_{0}=37.5$, $V_{\textrm{D}}=50.9$,
and $\sigma_{\textrm{D}}=0.25$ that correspond to
the regime $\xi<\sigma_{\textrm{D}}$ \cite{PhysRevA.82.033603}.

\begin{widetext}

\begin{figure}[tb]
\begin{center}
\includegraphics[height=3.28in,width=5.44in]{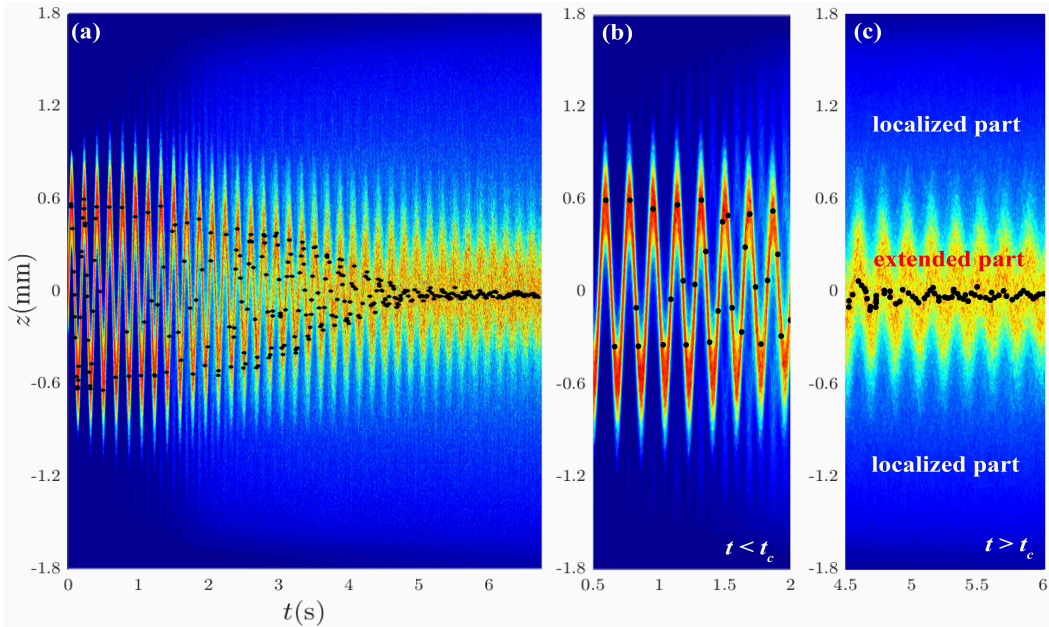}
\end{center}
\vspace{-0.6cm}
\caption{(a) Spatial and temporal distribution of an oscillating condensate.
The calculated norm of the spatial condensate
wave function are potted in $z$-direction at various times. Black dots
are the experimentally measured temporal center-of-mass coordinates, taken
from Fig.~2 of Ref.~\cite{PhysRevA.82.033603}.
(b) \& (c) Close examinations of the results for $t<t_c\approx 4$s (nonequilibrium)
and for $t>t_c$ (approaching the equilibrium in accompany of localization). }
\label{fig1}
\end{figure}

\end{widetext}

Fig.~\ref{fig1}(a) shows the spatial and temporal results of the oscillating condensate simulation.
We plot, at various times, the calculated norm of the spatial condensate
wave function in $z$-direction. For comparison, black dots
correspond to experimentally measured temporal center-of-mass coordinates
(reported in Fig.~2 of Ref.~\cite{PhysRevA.82.033603}).
Surprisingly it gives a very good agreement
between the simulation and the experiment.  Close examinations of the results are shown
in Fig.~\ref{fig1}(b) for $t<t_c\approx 4$s when the system is still out of equilibrium and
in Fig.~\ref{fig1}(c) for $t>t_c$ when the system is approaching the equilibrium.
In view of Fig.~\ref{fig1}(b), the calculated temporal density maxima match well
with the experimental data points.
One also sees that a minor (long-tail) part of atoms
oscillate out-of-phase to the major (central) part of atoms, which is consistent with
the experimental observation (see, for example, Fig.~5 in Ref.~\cite{PhysRevA.82.033603}).
In the experiment \cite{PhysRevA.82.033603}, the system was considered to be separated into a thermal
(non-condensed) component and a condensed component. At this earlier stage ($t<t_c$),
no major localization is yet to occur and our major (central) part corresponds to
their condensed component, while our minor (long-tail) part corresponds
to their thermal component.

Fig.~\ref{fig1}(c) shows the regime when the system is approaching the equilibrium
and the vast localization has occurred. In this regime,
the system is seen to consist of both localized and extended parts --
localized part mainly exists in the long-tail area and extended part mainly
exists in the central area \cite{PhysRevLett.106.040601}.
One sees that the oscillations of the central extended parts are significantly reduced
compared to those at $t<t_c$. It signals
that the system is approaching the equilibrium.
In contrast, the long-tail part is seen to be
completely static as the clear evidence of localization.
In the two-component scenario of Ref.~\cite{PhysRevA.82.033603}), it says that
the thermal (incoherent) component gets localized, while
the condensed (coherent) component remains extended.

\begin{figure}[tb]
\includegraphics[height=2.3in,width=3.3in]{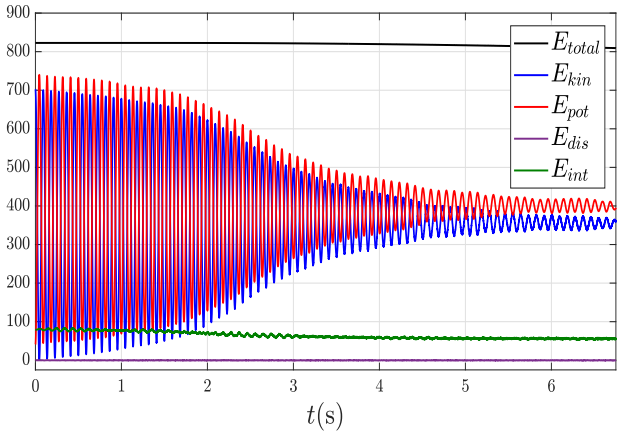}
\caption{Time evolution of the four energies in association with the dynamics shown in Fig.~\ref{fig1}.
Total energy is conserved for the entire process.}
\label{fig2}
\end{figure}

The good agreement between the simulation and the experiment
allows us to study in more details the two important phenomena,
{\em thermalization} and {\em localization}.
Total energy of the system consists of four terms:
$E_{\rm tot}(t)=E_{\rm kin}(t)+E_{\rm pot}(t)+E_{\rm dis}(t)+E_{\rm int}(t)$,
where the kinetic energy $E_{\rm kin}(t)=\int\left|\hbar\partial_{z}\psi\right|^{2}/\left(2m\right)dz$,
the potential energy $E_{\rm pot}(t)=\int V_{\textrm{ho}}\left|\psi\right|^{2}dz$,
the disorder energy $E_{\rm dis}(t)=\int V_{{\rm dis}}\left|\psi\right|^{2}dz$,
and the interaction energy $E_{\rm int}(t)=\left(Ng/2\right)\int\left|\psi\right|^{4}dz$.
Fig.~\ref{fig2} shows the time evolution of the four energies, respectively.
As random disorder potential is rapidly varying in space,
$E_{\rm dis}\simeq 0$ (purple line) for the entire process.
Moreover, because both the trapping and disorder potentials
are real and time-independent, one expects that there is no energy loss.
Conservation of total energy is indeed shown in Fig.~\ref{fig2} (black line).
Of most interest, during the process partial kinetic energy is in exchange
with partial potential energy and when $t\gg t_c$, both energies are expected to
come to a constant. It seems that the dissipation discussed in
Refs.~\cite{PhysRevA.82.033603,PhysRevA.82.053632,PhysRevLett.106.165301} can be realized as
the result of the exchange between kinetic and potential energies.

In the current weakly interacting system, $E_{\rm int}$ is relatively small compared to
both $E_{\rm kin}$ and $E_{\rm pot}$.
Thus the energy exchange occurs mainly
between $E_{\rm kin}$ and $E_{\rm pot}$. Moreover, as clearly seen in Fig.~\ref{fig2},
the energy exchange rate or the transportation of the condensate is
significantly reduced when $t> t_c$. One can also verify whether the virial theorem
is satisfied when the system approaches the equilibrium.
For the current system described by the GP equation (\ref{GPE}),
the following condition
\begin{equation}\label{virial}
  2E_{kin}-2E_{pot}+dE_{int}=0
\end{equation}
with $d$ the dimension should be satisfied at equilibrium \cite{RefB_Pitaevskii2016}.
From Fig.~\ref{fig2}, the condition $E_{pot}\simeq E_{kin}+E_{int}/2$ is indeed satisfied with $d=1$.

Owing to the randomness nature of the wave function,
it is particularly useful to study the corresponding waveaction spectrum $n_k(t)$ in the context of
wave turbulence \cite{wave_turbulence}.
When expressing the condensate wave function $\psi\left(z,t\right)$
in terms of Madelung transformation,
$\psi\left(z,t\right)=\sqrt{\rho\left(z,t\right)}\exp\left[i\varphi\left(z,t\right)\right]$
with $\rho$ and $\varphi$ the density and phase,
the hydrodynamic kinetic-energy density is
$\mathcal{K}=\left(m/2\right)|\left(\sqrt{\rho}\mathbf{u}\right)|^{2}$ with  $\mathbf{u}\equiv\left(\hbar/m\right)\partial_{z}\varphi$ the velocity.
To study the scaling laws, one applies the sum rule:
${\mathcal{K}}(t)=\int_{0}^{k_c}\widetilde{\mathcal{K}}\left(k,t\right)dk$,
where $\widetilde{\mathcal{K}}\left(k,t\right)$ is the kinetic-energy spectrum and
$k_c$ is the cutoff wavevector mentioned earlier.
The corresponding waveaction spectrum is then given by
$n_{k}\left(t\right)=k^{-2}\widetilde{\mathcal{K}}\left(k,t\right)$ and
one can define the entropy $S(t)$ in association with $n_k$
\cite{PhysRevLett.95.263901,Picozzi:07,10.1038/nphys2278,PhysRevX.7.011025},
\begin{equation}\label{entropy}
  S(t) =\int dk \ln\left[n_k(t)\right].
\end{equation}

\begin{figure}[tb]
\begin{center}
\includegraphics[height=2.3in,width=3.3in]{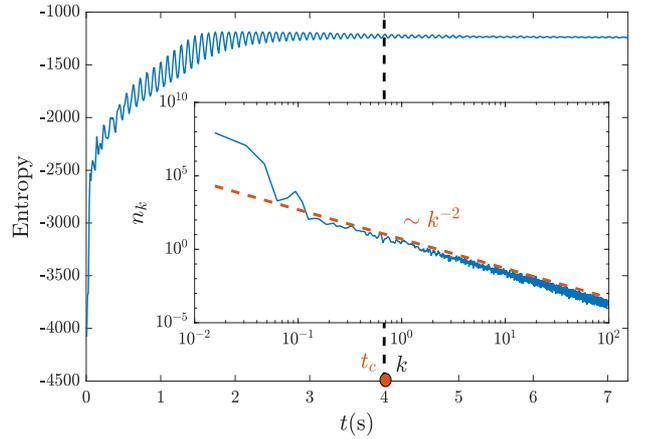}
\caption{Temporal entropy of the oscillating condensate calculated from the simulation in Fig.~1.
The inset shows the Rayleigh-Jeans spectrum for the waveaction
at $t=7.3$s when the system approaches the equilibrium.
$t_c\approx 4$s is the relaxation time.}
\label{fig3}
\end{center}
\end{figure}

Results of $S(t)$ are shown in Fig.~\ref{fig3}. Major features are that
$\Delta S(t)\equiv \lim_{\Delta t\rightarrow 0} [S(t+\Delta t)-S(t)]>0$
when $t < t_c$ and $\Delta S(t)\rightarrow 0$ when $t> t_c$. It unambiguously identifies that
thermalization is developed in the system.
In earlier process, one also sees the Fermi-Pasta-Ulam-Tsingou (FPUT) recurrence effect
that is consistent with the oscillation of the system.
In the inset of Fig.~\ref{fig3}, we show $n_k$ at $t = 7.3$s $\gg t_c$.
It follows the Rayleigh-Jeans spectrum, $n_k\sim k^{-2}$, which
indicates that the energy spectrum $\widetilde{\mathcal{K}}$
is a constant, or equipartition in $k$ space.
In other words, the system corresponds to a non-dissipative one
with a detailed balance.

Here we discuss the relaxation time $t_c$.
As studied by Bhongale {\em et al.} \cite{PhysRevA.82.053632},
by comparing the condensate center-of-mass speed $v$ to the
sound speed $c\equiv\sqrt{\mu/m}$,  the entire oscillation process can be divided into
fast or supersonic ($v>c$) and slow or subsonic ($v<c$) regions.
When $v\leq c$, the relatively slow motion of the (central) extended part
does not affect much the distribution of the (long-tail) localized part.
As a matter of fact, the relaxation time $t_c$ can be well defined as the time when
the center-of-mass speed is equal to the sound speed ($v=c$). It has been identified
for the experiment \cite{PhysRevA.82.033603} that $t_c\simeq 4$s when $v=c$ and it is
consistent with the value quoted according to the results of $S(t)$ in Fig.~\ref{fig3}.

\begin{figure}[tb]
\begin{center}
\includegraphics[height=1.5in,width=3.4in]{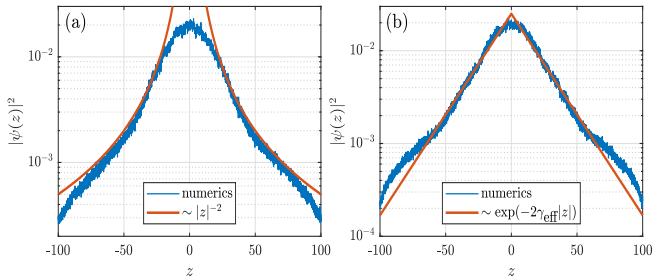}
\caption{(a) In a semi-log plot, for the case $\xi<\sigma_D$ the eventual density distribution
at $t=7.3$s is shown to exhibit an algebraic profile at $20<|z|<70$.
(b) A similar study for the case $\xi>\sigma_D$, where
an AL exponential profile is obtained for the eventual density distribution at $t=3.7$s.
The fitting Lyapunov exponent is $\gamma_{\textrm{eff}}=0.025$.}
\label{fig4}
\end{center}
\end{figure}

Finally we investigate in details whether a localization can be seen
in the oscillation experiment.
In a semi-log plot, Fig.~\ref{fig4}(a) shows the eventual spatial density distributions at $t=7.3$s.
To smoothen the spikes arising from randomness,
the results are taken as the average over a period.
As shown in Fig.~\ref{fig4}(a), it is verified that for the current $\xi<\sigma_D$ case
the condensate is algebraically localized, in accordance with
the previous theory \cite{PhysRevLett.98.210401}.
As mentioned earlier, in the presence of random
disorder, the condensate is actually separated into both localized and extended parts.
In a free expansion experiment, the extended part will escape whereas the
localized part will remain \cite{Nature.453.891}.
When a trapping potential is in place, the moving extended part
is eventually stopped at the center of the trap, whereas
the localized part comprises the long tails of the condensate.
Fig.~\ref{fig4}(a) shows a well-fitting curve
$|\psi(z)|^{2}\sim|z|^{-2}$
at the range $20<|z|<70$.
The lower bound is determined by the initial size of the condensate, {\rm i.e.}, the
Thomas-Fermi radius $R_{\textrm{TF}}=\sqrt{2\mu}=20$, and the off-fitting data out of the range
$|z|>70$ is due to the trapping effect.

Can the AL be seen in a similar oscillation experiment?
Here we perform another simulation for the same parameters except by
reducing $\sigma_{\textrm{D}}$ to 0.01 to make
the regime $\xi>\sigma_{\textrm{D}}$.
In this case, the relaxation time is $t_c\simeq 2$s only.
As shown in Fig.~\ref{fig4}(b), the results of
the eventual density distributions at $t=3.7$s are well fitted by an exponential one, $\left|\psi\left(z\right)\right|^{2}\sim\exp\left(-2\gamma_{\textrm{eff}}\left|z\right|\right)$
with $\gamma_{\textrm{eff}}$ the Lyapunov exponent \cite{PhysRevLett.98.210401}.
The fitting $\gamma_{\textrm{eff}}$
is also in good agreement with the analytic one,
$\gamma_{\textrm{eff}}=(\pi/32\xi)({V_{\rm D}/\mu})^{2}
(\sigma_{\textrm{D}}/\xi)\exp[-(\sigma_{\textrm{D}}/\xi)^{2}]\approx0.025$.
Compared to the case $\xi<\sigma_{\textrm{D}}$,
higher percentage of atoms can be localized in the case $\xi>\sigma_{\textrm{D}}$.
Thus one sees that the fitting is
as good as in the range $0<\left|z\right|<70$. The off-fitting data out of the range
is again due to the trapping effect.

In summary, we propose that Anderson localization can be observed in
an oscillating condensate in a disordered trap when the system comes to an equilibrium and
when the healing length of the condensate
is exceeding the disorder correlation length. In addition, we show that in such an ``isolated" system,
the disorder plays the role as a medium and through it, the system undergoes a
relaxation process from nonequilibrium to equilibrium.
The occurrence of localization can thus be viewed as the development of thermalization in
the system.

We are grateful to Randy Hulet for many useful comments.
Financial supports from MOST, Taiwan (grant No. MOST 105-2112-M-003-005),
JSPS KAKENHI (grant No. 17K05548) and MEXT KAKENHI/Fluctuation and Structure
(grant No.16H00807), and NCTS of Taiwan are acknowledged.

%

\end{document}